\documentclass[aps,prb,nopacs,floatfix,12pt]{revtex4} 

\usepackage{graphicx}
\usepackage{amssymb}
\usepackage{amsmath}
\usepackage{epsfig}
\usepackage{bm}
\usepackage{mathtools}
\usepackage{color}

\usepackage{rotating}

\begin{document}
\bibliographystyle{jcp}

\title{Infrared spectra of neutral bent carbon dioxide}
\author{Sergy\ Yu.\ Grebenshchikov\footnote{Email: Sergy.Grebenshchikov@ch.tum.de}}
\affiliation{Department of Chemistry, Technical University of Munich,
  Lichtenbergstr. 4, 85747 Garching, Germany} 

\begin{abstract}
A combined ab initio and quantum dynamical study characterizes a family of
bent neutral carbon dioxide molecules in terms of their vibrational levels,
electric dipole moment surfaces, and infrared spectra in the gas phase. 
The considered isomers include the dioxiranylidene (cyclic) form of 
CO$_2$ with the equilibrium valence angle of
72$^\circ$, belonging to the ground electronic state, and four open 
structures with the valence angles of 118$^\circ$/119$^\circ$
(belonging to the singlet and triplet electronic states 
$2^1\!A'$ and $1^3\!A'$, respectively) and 
127$^\circ$/128$^\circ$ (states 
$1^1\!A''$ and $1^3\!A''$, respectively). 
All studied bent structures possess permanent dipole moments. 
For all isomers, the antisymmetric
stretch fundamental is the strongest infrared transition. Individual bent
molecules can be distinguished on the basis of strong absorption bands in 
the frequency window 1100\,cm$^{-1}$ --- 1800\,cm$^{-1}$ as well as isotopic
shifts in the progression of antisymmetric stretch mode. 
Excitation of bent neutral carbon dioxide near 
a perfect metal surface is also briefly discussed. It is argued that 
the excitation energy from the linear ground state exhibits a red
red shift which depends on the molecule---metal distance. 
\end{abstract}

\maketitle

\section{Introduction}
\label{intro}

This paper explores the spectroscopic
properties of bent configurations of neutral CO$_2$ in the
low lying electronically excited singlet and triplet states with the 
equilibrium bending angles 
lying in the range of $70^\circ \le \alpha_{\rm OCO} \le 130^\circ$. The
focus is on the dipole moment functions and the vibrational infrared (IR) 
spectra. The analysis is performed using a combination 
of a high level electronic structure theory and quantum dynamics. 

Figure \ref{fig1}(a) gives an overview of the low energy portion of the
experimental vacuum ultraviolet (UV) absorption spectrum of CO$_2$ consisting
of two weak well separated bands.\cite{YESPIM96}  
In a recent ab initio quantum mechanical 
study, these bands were accurately reproduced  and the diffuse
absorption lines were assigned 
in terms of the principal vibrational modes of the molecule trapped in the
unstable resonance states 
[shown as a stick spectrum in panel (a)].\cite{G13B}  The
high energy band is mainly due to pseudorotational motion in the plane of
the two CO bonds, while the low energy band is dominated by the 
bending progressions.\cite{CLP92,G13B} These assignments are schematically
illustrated in Fig. \ref{fig1}(a). The present study originates from the 
observation that several lines near 65\,000\,cm$^{-1}$ stem from the 
\lq cyclic' OCO molecule, known as dioxiranylidene, with the valence angle of 
$\alpha_{\rm OCO} = 72^\circ$. The identification of this isomer
in the absorption spectrum 
has several chemically significant implications because bent CO$_2$
molecules are characterized by enhanced reactivity; one speaks of activated
carbon dioxide. 

Activation of carbon dioxide is often considered a
first step in the selective photoconversion 
of this inert greenhouse gas into  value-added chemicals, such as fuels.
The activation is commonly achieved photocatalytically, at a metal
or metal oxide surface acting as a photosensitizer.\cite{WDW15,KSIM15} 
The bent activated molecules in this case  are anion radicals 
CO$_2^{\gamma -}$ --- the key intermediates 
formed via a partial electron transfer from metal.\cite{FM86,B14A}
The negative electron affinity of carbon dioxide makes a 
single electron transfer to the molecule non-spontaneous. 
This is a limitation of the photocatalytic approach which affects its 
efficiency and forces one to look
for alternative activation and reduction schemes. 

One such scheme is the
activation/dissociation of the neutral CO$_2$ in the gas phase.
Preheating of the gaseous carbon dioxide gradually
shifts absorption from the vacuum UV range towards longer wavelengths 
at the UV edge of the solar spectrum making solar based activation 
feasible.\cite{TJ02,RVPG10} Recent measurements and calculations report 
appreciable absorption of preheated CO$_2$ at wavelengths reaching 
300\,nm.\cite{ODJH04,VFBGHLSDS13,G16A} The present study indicates that
near metal surfaces comparable red shifts in the absorption can be realized 
even at ambient temperatures. Plasmolysis
is another gas phase-based scheme of CO$_2$-to-fuel 
conversion.\cite{RBHMBBEGZS15,BKLS15,MZWAWT16} 
Bent CO$_2$ species near dissociation threshold are believed to affect the
inelastic electron excitation cross sections in plasma and to play the role
of transient species enhancing the 
dissociation rates and the carbon monoxide yields.

Bent states of CO$_2$ have a long history of studies. 
It is well known that its low lying excited states, 
both singlets ($2^1A'$ and $1^1A''$) and triplets 
($1^3A'$ and $1^3A''$), arising
from the excitations into the $6a_1$ component of the lowest unoccupied
$2\pi_u$ orbital, support bent equilibria as a direct consequence of the
Walsh 
rules.\cite{FM86,SFCCRWB92,BHLK00} Panel (b) of Fig.\ \ref{fig1} provides an
overview of  bent electronic 
states relevant for the present work. All states are 
labeled using the irreps of the $C_s$ symmetry group even though the actual
symmetry of the bent molecules is $C_{2v}$. 
The singlet and triplet states of $A'$ symmetry have equilibrium angles of
$118^\circ/119^\circ$; for the states of $A''$ symmetry, the 
 equilibrium angles are $128^\circ/129^\circ$.
For brevity, all bent equilibrium structures of carbon dioxide
will hereafter be termed \lq isomers'.
The OCO angle for the state $2^1A'$ (correlating at $C_{2v}$ 
geometries with $^1\!B_2$) was predicted by Dixon in his
seminal analysis of the CO flame emission bands.\cite{D63} Three decades
later, Cossart-Magos and co-workers\cite{CLP92} detected 
two bending progressions in the wavelength range 172\,nm---198\,nm
(photon energies between 51\,000\,cm$^{-1}$ and
57\,000\,cm$^{-1}$) which they assigned to the bent state $1^1\!A''$
excited in a perpendicular transition from $\tilde{X}$. 
The near-equilibrium shapes of the potential energy functions of the bent
singlet and triplet states were 
investigated by Spielfiedel et al.\cite{SFCCRWB92} at the CASSCF level
of theory (complete active space self-consistent field method). Recently, the
global potential energy surfaces of the six lowest singlet states of CO$_2$
were mapped out using the internally-contracted 
multireference configuration interaction singles and doubles (MRD-CI) method 
and a large atomic basis set.\cite{G13A}
 These calculations accurately 
reproduced the measured absorption spectrum and allowed definitive assignments
of many diffuse spectral bands.\cite{G13B} 
The UV absorption spectra of the 
bent triplet states were calculated by Schmidt et al.,\cite{SJSCH13}
while the singlet-singlet and 
triplet-singlet emission spectra were recently analyzed in Ref.\ 
\onlinecite{PG16B}. 

Despite numerous studies, a systematic spectroscopic analysis of the 
neutral 
bent isomers remains outstanding and their vibrational states are poorly
understood. A case in point is dioxiranylidene, which was 
predicted by Feller et al.\cite{FKD80} 
Its equilibrium geometry and the energetic stability were analyzed in
Refs. \onlinecite{XR94} and \onlinecite{HM00A}. While its signature
in the vacuum UV absorption spectrum has been identified,\cite{G13B} 
the vibrational spectrum of cyclic CO$_2$ is practically unknown. 

In this study I extend the previous work and present a detailed 
analysis of the dipole moment functions and the IR 
vibrational spectra of the five stable bent isomers 
of CO$_2$ in the  low lying 
excited singlet and triplet electronic states.  It is hoped
that the predicted IR intensities and the isotope 
shifts, calculated for several isotopomers,  can help experimental
identification of the neutral activated carbon dioxide intermediates. 

The paper is organized as follows:  The ab initio calculations and the 
potential energy profiles along the bending angle 
are discussed in  Sect.\ \ref{pes} together with the  corresponding 
dipole moment functions. Section \ref{vib} briefly characterizes the  
vibrational states of the bent isomers, compares their fundamental
frequencies with the known values for 
the linear CO$_2$ and for the bent anion radical 
CO$_2^-$, and discusses the IR spectra 
for excitations out of the ground vibrational states in the bent
equilibria. The isotope dependence of the IR spectra is
illustrated for the singlet
electronic states. Section \ref{conc} concludes and discusses a possible
extension of this study to include stabilization of the polar bent isomers
near metal surfaces due to dipole-image dipole interactions.

\section{Static properties of bent isomers of CO$_2$}
\label{pes}

Three-dimensional potential energy surfaces (PESs) of the ground electronic 
state $\tilde{X}^1\!A'$ and the bent singlet states 
$2\!A'$ and $1\!A''$ have been described in detail in Ref.\
\onlinecite{G13A}. They were calculated using
 the MRD-CI method  based on the CASSCF calculations
with 16 electrons in 12 active orbitals and 6 electrons
in three fully optimized closed-shell inner orbitals.
The doubly augmented correlation consistent polarized valence quadrupole
zeta (d-aug-cc-pVQZ) atomic basis set due to Dunning,\cite{D89} is used.
All calculations were performed using the MOLPRO package.\cite{MOLPRO-FULL}
The quality of  the MRD-CI potential energy surfaces in the Franck-Condon
(FC) zone has been discussed in Refs.\ \onlinecite{G13B} and 
\onlinecite{G16A}. In the interaction region, the calculated PESs agree well
with the  potentials recently 
calculated  at the coupled cluster (EOM-CCSD) level of theory.\cite{ASSG17} 

Here, the calculations of Ref. \onlinecite{G13B} are 
improved in one respect: New points are added to the angular grid  
near $\alpha_{\rm OCO} = 72^\circ$, $118^\circ$, and $127^\circ$, 
as well as in the range $30\le \alpha_{\rm OCO} \le 60^\circ$
($\sim 500$
geometries in total are additionally calculated) in order to improve
the description of bent equilibria. Their characteristic features 
are summarized in Table \ref{table1}; the potential cuts along the 
minimum energy path are shown in Fig.\ 
\ref{fig1}(b). The cyclic isomer CO$_2(72^\circ|\tilde{X}^1\!A')$ 
in the adiabatic ground state $\tilde{X}^1\!A'$ is C$_{2v}$ symmetric, with 
two CO bond lengths of 2,51$\,a_0$ substantially elongated relative to the 
linear equilibrium of CO$_2(180^\circ|\tilde{X}^1\!A')$. 
It is separated from the global
linear minimum by a $\sim 0.45$\,eV high potential barrier at 
$\alpha_{\rm OCO} = 90^\circ$.\cite{NOTE-CO25-1} 
The electronic origin of 
the cyclic minimum can be traced to the state $4^1\!A'$ [not shown in Fig.
\ref{fig1}(b)] which at linear geometries becomes a component of the 
orbitally degenerate state 
$2^1\Delta_u'$ resulting from an electronic promotion from the molecular 
orbital $1\pi_u$ into the Rydberg-type orbital 
$3p\pi_u$.\cite{SFCCRWB92,BHLK00,G13A} The other two singlet isomers
CO$_2(118^\circ|2^1\!A')$  and CO$_2(127^\circ|1^1\!A'')$ are also
C$_{2v}$ symmetric (see Table \ref{table1}). These open structures have
similar energies and are located $\sim 0.5$\,eV below the cyclic minimum.
All three bent equilibria in the singlet states lie below 
the threshold of the
singlet dissociation channel ${\rm CO}(\tilde{X}^1\Sigma^+) + {\rm O}(^1D)$.
The two open structures 
are also quasi-stable relative to the lower energy triplet dissociation channel
${\rm CO}(\tilde{X}^1\Sigma^+) + {\rm O}(^3P)$ (the potential minima 
practically coincide with the threshold energy). Note however that the  
spin-orbit coupling in CO$_2$ 
is weak, of the order of 80\,cm$^{-1}$,\cite{ASSG17} 
and the rate of internal conversion
into the triplet dissociation channel is expected to be small.

The triplet states $1^3\!A'$  and $1^3\!A''$ are calculated with the same
atomic basis and at the same level of theory as the singlets. The calculations
are performed on a three dimensional grid of internal molecular coordinates,
with the CO bond distances ranging from $1.8\,a_0$ to $5.0\,a_0$, and the 
valence angle in the range $\alpha_{\rm OCO} = [70^\circ,175^\circ]$; energies
for $\alpha_{\rm OCO} < 70^\circ$ are extrapolated.  
The triplet potentials along the bending angle are compared with the
singlets in Fig.\ \ref{fig1}(b). The states $1^3A'$ and  $1^3A''$ have minima
located at nearly the same angles as their singlet counterparts
$2^1A'$ and  $1^1A''$ [cf. Table \ref{table1}].
In each singlet/triplet pair of open isomers, the triplet one is more 
stable (by 0.50\,eV for $2^1A'/1^3A'$ and by 0.19\,eV for 
$1^1A''/1^3A''$). Both triplet states lie below the asymptote 
${\rm CO}(\tilde{X}^1\Sigma^+) + {\rm O}(^3P)$. 

The ab initio electric dipole moments in different electronic states are 
computed with MOLPRO at the MRD-CI level of theory 
using the same coordinate grid as for the PESs above. 
The molecular axes $(x',y',z')$ 
in these calculations are chosen as in Ref.\ \onlinecite{G13A}, i.e. 
$x'$ is orthogonal to the molecular plane, $z'$  runs
along one of the CO bonds and $y' \perp z'$. The in-plane components of the 
electric dipole are generally non-zero, while 
the component along $x'$ vanishes. For further analysis, the 
 dipole moments are transformed to the bisector frame $(x,y,z)$ , with
$x = x'$, the axis $z$ lying on
a bisector of the $\alpha_{\rm OCO}$ angle, and $y \perp z$.  Values between 
the grid points are obtained by interpolation using cubic splines. 

Cuts through the dipole moment surfaces $\mu_{y,z}(R_1,R_2,\alpha_{\rm OCO})$ 
in the plane $(R_1,R_2)$ of the two CO bond lengths
are shown in Fig.\ \ref{ir01}(a,b,c) for the singlet states, and 
in Fig.\ \ref{ir02}(a,b) for the triplet states. The bending angle is fixed
to the equilibrium value for the given bent state. The components of the 
dipole moments obey 
simple symmetry rules with respect to the interchange of the two CO 
bonds:
\begin{eqnarray}
\label{symu}
\mu_z(R_1,R_2) & = & \mu_z(R_2,R_1) \nonumber \\
\mu_y(R_1,R_2) & = & -\mu_y(R_2,R_1) \, ;
\end{eqnarray}
the component $\mu_y$ (irrep $b_2$) is antisymmetric  and vanishes 
in C$_{2v}$-symmetric equilibria, 
while the component $\mu_z$ is symmetric (irrep $a_1$), so that 
the bent isomers
 possess permanent electric dipoles. The dipole moments at equilibrium,  
$\mu_z \equiv \mu_{\rm eq}$, are given in Table \ref{table1}, and the 
corresponding dipole vectors are illustrated in Fig.\ \ref{fig1}(b). 
The dependence of $\mu_{\rm eq}$ on the equilibrium 
bending angle is non-monotonic, with the largest dipole moment
of 1\,D found for 
$\alpha_{\rm OCO} = 118^\circ$ for the singlet state $2^1\!A'$. The
cyclic CO$_2$ has a dipole moment of -0.25\,D. This is the only 
isomer, for which the direction of the
$\mu_z$ vector, $\rm C^{\delta-} \rightarrow O_2^{\delta}$,  
is opposite to what one would expect on the basis of the electronegativity. 
The sign reversion of the dipole moment function was studied by Harrison for
carbon monoxide.\cite{H06} He demonstrated that the net sign of the dipole
moment is determined by a trade off of the direct charge contribution and
the induced atomic dipoles. The 
permanent dipoles of the triplet states are generally smaller than those of
their twin singlet states, and the dipoles of
all neutral bent isomers are comparable or larger than the dipole  
of the carbon dioxide anion CO$_2^-$.

While the permanent dipole moments characterize bent equilibria, 
the dependences of $\mu_y$ and $\mu_z$ on the displacements from the 
equilibrium geometry control the intensity patterns in
the vibrational IR spectra. The reflection
symmetry of the components, Eq.\ (\ref{symu}), is utilized in 
 Figs.\ \ref{ir01}(a,b,c) and \ref{ir02}(a,b): In each panel, the antisymmetric
function $\mu_y$ is shown above and the symmetric function $\mu_z$ is shown 
below the diagonal $R_1 = R_2$.
Near the diagonal, the component $\mu_z$ grows linearly 
along the symmetric stretch coordinate $R_+ = (R_1+R_2)/2$ and along  
the bending angle $\alpha_{\rm OCO}$; the component
$\mu_y$ grows linearly  along the antisymmetric 
 stretch coordinate $R_- = (R_1-R_2)/2$. The slopes of these linear
dependences are the coefficients of the Herzberg-Teller 
expansion for the dipole moments;\cite{BUNKERJENSEN06} 
their absolute values 
are listed in Table \ref{table1}.  Although $\mu_z \gg \mu_y$ for
both singlet and triplet states in the FC zone, the weak component 
$\mu_y$ grows rapidly with $R_-$ implying that an intense line of the
 antisymmetric stretch fundamental can be expected in the IR spectra. For
two singlet isomers, CO$_2(72^\circ|X^1\!A')$ and CO$_2(118^\circ|2^1\!A')$,
the components $\mu_y$ are monotonic functions of $R_-$. In
contrast, the functions  $\mu_y$ for the isomers 
CO$_2(127^\circ|X^1\!A'')$, CO$_2(119^\circ|1^3\!A')$, and
CO$_2(128^\circ|1^3\!A'')$ develop pronounced maxima close to the diagonal line
$R_1 = R_2$. For these isomers, higher derivatives of $\mu_y$ are expected
to significantly contribute to the Herzberg-Teller expansion.  

\section{The infrared spectra of bent isomers of CO$_2$}
\label{vib}

The vibrational states of bent CO$_2$ isomers are calculated quantum 
mechanically using the program package \lq PolyWave'\cite{NOTE-CO21-4}  (see
Appendix for the computational details). 

The fundamental frequencies of the symmetric stretch ($\omega_s$), 
bend ($\omega_b$),  and antisymmetric stretch ($\omega_a$) 
vibrations are listed
in  Table \ref{table1} and compared with the known
frequencies of the linear CO$_2$ and of the CO$_2^-$ anion radical. 
The frequencies $\omega_s$ and $\omega_b$ 
only weakly depend on the equilibrium angle. 
The calculated bending frequencies lie
between between 600\,cm$^{-1}$ and
700\,cm$^{-1}$; the symmetric stretch frequencies range from about 
1300\,cm$^{-1}$ to 1500\,cm$^{-1}$ (an exception is the isomer 
CO$_2(128^\circ|1^3\!A'')$ in which $\omega_s = 1088$\,cm$^{-1}$ is 
rather low). However, the 
deviations relative to the linear molecule are
sufficient to tune the frequencies  $\omega_s$ and $\omega_b$
away from the anharmonic
$1:2$ Fermi resonance\cite{F31}  which dominates the vibrational spectrum
of linear OCO. The calculations agree well with the
experimental estimates\cite{CLP92} 
 available for the state $1^1\!A''$  
($\omega_b \approx 600$\,cm$^{-1}$ and $\omega_s \approx
1370$\,cm$^{-1}$), as well as with the previous theoretical results. 
The frequency of the antisymmetric stretch mode undergoes the 
most conspicuous change and in the bent isomers 
substantially deviates from $\omega_a = 2349$\,cm$^{-1}$ of
the linear CO$_2(180^\circ|X^1\!A')$. The authors of 
Ref.\ \onlinecite{SFCCRWB92}
came to the same conclusion using CASSCF method. 
The antisymmetric stretch frequency
tends to decrease with decreasing equilibrium bending angle reaching the
 value of 
683\,cm$^{-1}$ in the cyclic CO$_2(72^\circ|X^1\!A')$); in fact, 
$\omega_a$  and $\omega_b$ in this case 
become accidentally degenerate. Table \ref{table1}
demonstrates that the dependence of $\omega_a$ on 
$\alpha_{\rm OCO}$ is not simple and not monotonous. 
In particular, the antisymmetric stretch frequencies
of the triplet states are different from those of the singlet states with
the same equilibrium angle. Nevertheless, 
the frequency $\omega_a$ remains the most 
convenient proxy for the geometry of the activated CO$_2$. As discussed below, 
the antisymmetric stretch fundamental is
the strongest transition in the IR spectra of bent CO$_2$.


Anharmonic couplings
between vibrational modes in bent isomers are generally weak. 
Many vibrational eigenfunctions $\psi_{\bm n}$
have unperturbed nodal lines and
can be assigned quantum numbers $v_s$, $v_b$, and $v_a$ corresponding to
the symmetric stretch, bend,
and antisymmetric stretch modes, respectively (${\bm n} = (v_s,v_b,v_a)$ is a  
vector index). Table \ref{table2} in the Appendix
summarizes the assigned vibrational states
with energies up to 5000 cm$^{-1}$ above the ground vibrational level. 
The potential wells for different isomers have 
different depths (see Table \ref{table1}) and support
different numbers of bound states. The longest lists of assignable states 
are found for the bent molecules
CO$_2(118^\circ|2^1\!A')$, CO$_2(127^\circ|1^1\!A'')$, and 
 CO$_2(119^\circ|1^3\!A')$; for them, the potential wells are deep, 2\,eV for
the two singlet species, and 1.2\,eV for the triplet one. 
The potential wells of the remaining two isomers are rather shallow. 
 For CO$_2(72^\circ|\tilde{X}^1\!A')$, only the vibrational levels with 
energies below the $\sim 0.45$\,eV high potential barrier, separating 
the cyclic minimum from the region of $\alpha_{\rm OCO} > 90^\circ$, 
are included (states localized in the cyclic well and 
belonging to the pure bending progression can also be found at higher 
energies, even 1.5\,eV 
above the barrier top\cite{G13B}). The well of the triplet
carbon dioxide CO$_2(128^\circ|1^3\!A'')$ is separated from the dissociation
channel by a potential barrier which in the present MRD-CI calculations lies
0.56\,eV above minimum (the  authors of 
of Ref.\ \onlinecite{SFCCRWB92} reported this state, calculated using CASSCF 
method, as unbound). 
Note however, that only  short vibrational progressions are found 
in the IR spectra discussed below,  and  even moderately 
excited vibrational states with the quantum numbers larger than 
3 or 4 do not show up in the spectra.

For each isomer, the IR intensities $I_{\bm n}$ for transitions out of the 
ground vibrational state $\psi_0$ are evaluated using 
the matrix elements involving the precalculated 
vibrational wave functions $\psi_{\bm n}$ and the 
 components $\mu_{y,z}$ of the dipole moment: 
\begin{equation}
\label{irdmu}
I_{\bm n}
\sim E_{\bm n}\left[\left|\langle \psi_0 |\mu_y| \psi_{\bm n}\rangle\right|^2
+\left|\langle \psi_0 |\mu_z| \psi_{\bm n}\rangle\right|^2\right] \, .
\end{equation}
Here $E_{\bm n}$ is the vibrational energy relative to the ground state. 
For C$_{2v}$ symmetric isomers  with 
the isotopic composition $^{12}$C$^{16}$O$_2$, 
only one of the two matrix elements in Eq.\ (\ref{irdmu}) is 
non-zero for a given ${\bm n}$. Indeed, the dipole moment function 
$\mu_y$ is of $b_2$ symmetry and promotes excitations to 
states with odd number of antisymmetric stretch quanta $v_a$; the totally
symmetric function
$\mu_z$ mediates transitions to the complementary set of states with even
$v_a$ values.

The calculated IR spectra for the singlet bent isomers
are shown in Fig.\ \ref{ir01}(d,e,f); the IR spectra of the triplet species
are in Fig.\ \ref{ir02}(c,d). 
The intensities $I_{\bm n}$ are given relative to the intensity $I_{010}$ 
of the fundamental bending excitation $(0,1,0)$. 
The IR spectra of all bent
isomers are dominated by the transition to the antisymmetric stretch
fundamental $(0,0,1)$. This can be explained by comparing the 
Herzberg-Teller expansion coefficients of
different vibrational modes in Table \ref{table1}.   
The gradient $\partial \mu_y/\partial R_-$ of the dipole
moment $\mu_y$ is 
noticeably larger than the gradients of $\mu_z$ with respect to 
$\alpha_{\rm OCO}$ or  $R_+$. The relative intensities
for the antisymmetric and symmetric stretch 
fundamentals $I_{001}/I_{010}$ 
and  $I_{100}/I_{010}$  are also listed in Table \ref{table1}. 
For CO$_2(72^\circ|\tilde{X}^1\!A')$, the intensity $I_{001}/I_{010}\sim 9$ 
is the smallest; it is close to the experimental value of 12 for 
the linear isomer CO$_2(180^\circ|X^1\!A')$.\cite{RY81} For all other
isomers in Table \ref{table1}, this intensity is noticeably
larger and 
is either comparable to the value of $I_{001}/I_{010} = 28$ 
predicted\cite{ZA99}
for the anion radical CO$_2^-$ or exceeds it.  The largest 
intensity of $I_{001}/I_{010}  = 107$ is found for the singlet
isomer CO$_2(118^\circ|2^1\!A')$. The different relative intensities 
$I_{001}/I_{010}$ 
calculated for different isomers are primarily due to the variations in 
the intensity of the bending fundamental, which in turn correlate with the 
Herzberg-Teller coefficients
$\partial \mu_z/\partial \alpha_{\rm OCO}$. The actual vibrational wave 
functions are anharmonic, and the correlation is often imperfect (see
Table \ref{table1}). In all bent molecules, the symmetric stretch mode is
IR active. However, the relative intensity is modest, 
$I_{100}/I_{010}  \sim 1$, 
matching that predicted for CO$_2^-$.\cite{ZA99} The exception is 
CO$_2(72^\circ|\tilde{X}^1\!A')$ for which this line is
rather intense, $I_{100}/I_{010}  = 7$, 
and can be used as a spectral marker of dioxiranylidene.  

Despite many common features, the IR intensity patterns form a 
unique signature of a given isomer. 
For CO$_2(72^\circ|X^1\!A')$, the IR peaks are resolved only within 
2000\,cm$^{-1}$ above the ground state [Fig.\ \ref{ir01}(d)]
--- the coordinate dependences of
the dipole moments are close to linear. The frequency
window 1100\,cm$^{-1}$ --- 1800\,cm$^{-1}$, often used for the assignment
of bent isomers,\cite{IKS11} contains two excitations
lying 226\,cm$^{-1}$ apart,  the
already mentioned intense line $I_{100}$ and the second overtone of the 
antisymmetric stretch mode $I_{002}$, both promoted by the dipole moment
$\mu_z$.  

The IR spectrum of CO$_2(118^\circ|2^1\!A')$ extends over the energy
range of more than 4000\,cm$^{-1}$ [Fig.\ \ref{ir01}(e)]. The main transitions
stem from the dipole moment $\mu_y$ and terminate on the vibrational states 
with one quantum of antisymmetric stretch $v_a = 1$. 
Clearly seen in the spectrum is the bending progression $(0,v_b,1)$ with
$v_b \le 5$ and the intensity which gradually decreases with increasing
$v_b$.  The frequency window
1100\,cm$^{-1}$ --- 1800\,cm$^{-1}$ contains only one line $I_{011}$; it lies
566\,cm$^{-1}$ above $I_{001}$. The spectrum of the triplet incarnation of 
this isomer, CO$_2(119^\circ|1^3\!A')$, is in many respects similar. 
The bending progression, however, is short while the 
symmetric stretch excitations with $v_s  = 1$ are clearly seen
[Fig.\ \ref{ir02}(c)]. The frequency window
1100\,cm$^{-1}$ --- 1800\,cm$^{-1}$ contains the stronger line $I_{011}$ and a 
weak peak of $I_{100}$. The asymmetric stretch line $I_{001}$ lies just a
few cm$^{-1}$ below this window.

Characteristic for IR spectrum 
of the singlet isomer CO$_2(127^\circ|1^1\!A'')$ 
are excitations of the antisymmetric stretch overtones with
$v_a = 1, 3$, as well as an unusually long progression of the symmetric
stretch excitations with $v_s \le 3$ [Fig.\ \ref{ir01}(f)]. 
Excitations, falling into  
the frequency window 1100\,cm$^{-1}$ --- 1800\,cm$^{-1}$, include two peaks, 
$I_{100}$ and $I_{011}$, lying 293\,cm$^{-1}$ apart.  The spectrum of the
triplet isomer CO$_2(128^\circ|1^3\!A'')$ also includes transitions to
states with $v_a = 1, 3$. The 
anharmonic couplings in this isomer are unusually 
strong, and several peaks in 
the vibrational spectrum are difficult to assign uniquely. The frequency
window 1100\,cm$^{-1}$ --- 1800\,cm$^{-1}$ includes a single assigned peak 
$I_{101}$ lying $\sim 1000$\,cm$^{-1}$ above $I_{001}$. 


Shifts in the positions of the IR absorption lines upon isotope 
substitution often facilitate the experimental assignment, and 
isotope shifts in vibrational progressions are considered to be sensitive 
indicators of the isomer geometry. For example for CO$_2^-$, the 
measured isotope shifts were used to estimate the equilibrium angle of the
radical.\cite{ARWCM69,GE81} In this work, the isotope shifts are calculated
for the singlet isotopomers, two of which are $C_{2v}$
symmetric,  
$\rm ^{13}C^{16}O_2$ and $\rm ^{12}C^{18}O_2$, and one has the lower
$C_s$ symmetry, $\rm ^{16}O^{12}C^{18}O$. They are abbreviated as 636, 828,
and 628, respectively; in the first, the central 
carbon atom is substituted, while in the other two the end oxygen atoms are
substituted. Their IR spectra are calculated
as described in Appendix, and the isotope shifts are found by matching 
states with the same vibrational assignments. The largest isotope shifts
are found for the bending states $(0,v_b,0)$ and the antisymmetric stretch
states $(0,0,v_a)$. They are shown in Fig.\ \ref{isofig}. 
Different bent isomers are marked with different
colors, and different isotope combinations 
are marked with different symbols; the
vibrational states are depicted regardless of their IR intensity. 
Common to most calculated isotope shifts is their near linear growth inside
progressions with increasing vibrational energy. This indicates that the 
progressions are approximately harmonic. Indeed, for a harmonic oscillator 
the dependence of the isotope shift $\delta G$ 
on the vibrational quantum number $v_n$ (or the vibrational energy 
${\epsilon_n}$) is given by 
\begin{equation}
\label{iso01}
\delta G(v_n) = \left(1-\frac{\omega_n^i}{\omega_n^0}\right)
\hbar\omega_n^0(v_n + 1/2) \equiv (1-\rho_n)\epsilon_n
\end{equation}
with the slope determined by the ratio $\rho_n$
of the harmonic frequencies 
in the unsubstituted ($\omega_n^0$) and substituted ($\omega_n^i$) molecules.
Dashed lines in Fig.\ \ref{isofig} depict the corresponding predictions 
for each progression and each isotopomer. 

For the bending progression
(left panel), the harmonic prediction is rather accurate. The largest shifts
are observed in the 828 isotopomer, while 636 gives the lower bound. 
As expected, the
bending progression is characterized by large oxygen and 
small carbon isotope shifts. Note that for a given isotopomer, the symbols
corresponding to different bent equilibria (and therefore
differently colored) cluster around one straight line: 
The isotope shifts in the bending progression 
are not sensitive to the equilibrium
$\alpha_{\rm OCO}$ angle and are similar for all bent isomers.

The isotope shifts in the antisymmetric stretch progression follow a 
different pattern (right panel). The smallest shifts are in the 828 
isotopomer, and the largest shifts are found for 636, 
indicating that this progression responds stronger to the carbon
substitution.  The harmonic approximation appears accurate only for the
open isomers CO$_2(118^\circ|2^1\!A')$ and CO$_2(127^\circ|1^1\!A'')$
(red and green symbols). For the cyclic isomer CO$_2(72^\circ|X^1\!A')$
(blue symbols), the deviations from the predictions of 
Eq.\ (\ref{iso01}) are strong. 
For example in the cyclic 636 isomer, the slope of 
the dependence $\delta G$ vs. $\epsilon_n$ is smaller than the 
prediction by a factor or two. 
In the cyclic 828 and 628 molecules, 
the actual slopes a larger than predicted. 
As a result, the isotope shift in the oxygen substituted cyclic
828 isotopomer is larger than in 
the carbon substituted cyclic 636 isotopomer, 
contrary to the expectations. Thus, the 
isotope shifts in the antisymmetric stretch progression $(0,0,v_a)$
allow one to discriminate between the dioxiranylidene structure
CO$_2(72^\circ|X^1\!A')$ and the open configurations 
CO$_2(118^\circ|2^1\!A')$/CO$_2(127^\circ|1^1\!A'')$. 


\section{Conclusions and outlook}
\label{conc}

This paper analyzes the properties of the activated neutral 
carbon dioxide molecules CO$_2(72^\circ|\tilde{X}^1\!A')$, 
CO$_2(118^\circ|2^1\!A')$,  CO$_2(119^\circ|1^3\!A')$, 
CO$_2(127^\circ|1^1\!A'')$, and CO$_2(128^\circ|1^3\!A'')$. 
Their IR spectra are calculated using the new ab 
initio potential energy and electric dipole moment surfaces of the
bent triplet states $1^3\!A'$ and $1^3\!A''$, 
as well as the previously reported (and improved)
surfaces for the bent singlet states $2^1\!A'$ and $1^1\!A''$
and for the local dioxiranylidene minimum of the ground electronic
state $\tilde{X}^1\!A'$. The main results can be summarized as follows:
\begin{enumerate}
\item The antisymmetric stretch fundamental 
$(0,0,1)$ is the strongest transition out of the ground vibrational state
for all bent isomers. The intensity of this transition, 
$I_{001}/I_{010}$, measured relative to the intensity of the bending 
fundamental, is much larger than in linear CO$_2$ for all bent molecules but  
the triplet isomer CO$_2(128^\circ|1^3\!A'')$. 
\item Individual isomers can be distinguished on the basis of 
their strong IR bands in the broad frequency interval,
shown in Figs.\ \ref{ir01} and \ref{ir02}, as well as in the narrow 
frequency
window 1100\,cm$^{-1}$ --- 1800\,cm$^{-1}$. A unique
feature of the cyclic isomer CO$_2(72^\circ|\tilde{X}^1\!A')$ in this window
is a strong excitation of the symmetric stretch fundamental. 
\item Isotope shifts, calculated for three singlet isotopomers,  
can be used as fingerprints of activated neutral CO$_2$ molecules. 
The shifts grow in the pure bending and antisymmetric stretch 
progressions, reaching $\ge 40$\,cm$^{-1}$ already for the second pure 
overtones. 
For the cyclic CO$_2$, the calculated isotope shifts strongly deviate from 
the nearly harmonic shifts predicted for other bent isomers. 
\end{enumerate}


The bent neutral isomers in the gas phase 
are generated via an electronic excitation
across the large HOMO-LUMO gap. This requires the energy of 
at least 4.5\,eV (for the
triplet state) or 5.4\,eV (for the singlet state) to be supplied by the
incident light; the corresponding wavelengths are 274\,nm and 230\,nm,
respectively. One way to reduce the excitation energy 
is to heat CO$_2$ gas and to irradiate
vibrationally excited parent molecules.\cite{ODJH04,RVPG10,G16A} 
Temperatures over 2500\,K are
needed to substantially shift the optical 
absorption to wavelengths larger than 250\,nm.\cite{ODJH04,RVPG10,G16A} 

The gas phase data discussed in this paper can be useful in exploring and
designing alternative routes to activate neutral CO$_2$.
Table \ref{table1} demonstrates that all bent isomers are polar molecules
with permanent dipole moments. This suggests that the spectroscopy
of CO$_2$ is sensitive to the environment effects.  
For example, interaction with
a metal surface lowers the energy of a dipole, so that the optical excitation
gap from the nonpolar linear ground electronic state into polar bent states
becomes a function of the metal---molecule distance. The simplest estimate
of this dependence is provided by classical electrostatics. The change in
the energy $E^\star$
of a polar excited state due to the interaction of a classical dipole
$\mu$ with its image in a perfect metal surface is given by\cite{HH86}
\begin{equation}
\label{diposhift}
E^\star(z) = E^\star_\infty - \frac{\mu^2}{8(z-z_{\rm im})^3} \, .
\end{equation}
Here $E^\star_\infty$ is the gas phase energy of
the bent excited state relative to the nonpolar ground electronic state;
$z$ is the distance from the center of mass of CO$_2$ to the metal
image plane located at $z_{\rm im}$; the dipole is assumed to lie 
perpendicular to the metal surface (carbon-down arrangement). 
CO$_2$ physisorbs on most metals,\cite{B14A} and the Eq.\ (\ref{diposhift}) 
is applied to Cu(100) for which the image plane is located
at $z_{\rm im} = 2.3\,a_0$ in front of the last Cu plane.\cite{BGKCSE01} 
Figure \ref{meco2} shows the distance dependent excitation energies for
CO$_2(118^\circ|1^1\!A')$ with the dipole moment of $\mu = 0.99$\,D and
for CO$_2(119^\circ|1^3\!A')$ with the dipole moment of $\mu = 0.46$\,D. 
The influence of the image potential is appreciable, and 
1 bohr away
from the image plane, i.e. $3.5\,a_0$ away from the metal surface, 
the original excitation energy of the singlet state is
decreased by almost 1\,eV.  
A comparable reduction in the excitation energy
can be achieved in the homogeneous gas phase only 
at temperatures over 7000\,K. For the triplet state with smaller dipole moment,
the excitation energy shifts by about 0.25\,eV. 
Incidentally, these red shifts imply that the polar states stabilize
relative to the dissociation thresholds because carbon monoxide 
CO$(\tilde{X}^1\Sigma^+)$ 
has a very small equilibrium electric dipole and is not
affected by image forces.

Although the electrostatic effects are expected to be substantial at distances 
over $3\,a_0$ from a metal surface,\cite{GLSV12,FKPH09} the actual 
CO$_2$/metal interaction is complicated, and the above estimate serves only
for orientation. Interaction with metal surface significantly affects the
electronic structure of the adsorbate molecule. Even in the simple model
of Fig.\ \ref{meco2}, the singlet and the triplet states are seen to cross
at distances $z - z_{\rm im} < 0.9\,a_0$. Quantum chemical calculations are
therefore mandatory for a quantitative analysis of the 
optical excitation gap at
different absorption sites and the degree of spin quenching in the
electronically excited states near a metal surface. First principles
calculations of the molecular excited electronic
states near metal surfaces is a challenging problem,\cite{KGWC01} and this
is the direction in which this study will be extended. The planned calculations
will also allow one to address 
questions, related to the kinetics of the excited states, and to 
estimate the characteristic 
rates of electron transfer into metal, the rates of radiationless
relaxation into the ground electronic state, and to compare them with the
desorption rates of electronically excited CO$_2$. 


\appendix

\section{Quantum mechanical calculations of the vibrational spectra}
\label{appa}

Bound
vibrational states in the potential wells of the bent singlet and 
triplet isomers 
are found by applying Filter Diagonalization\cite{MGT95} to the
respective Hamiltonians set in the Jacobi coordinates $R$, $r$, and $\gamma$.
Here $R$ is the distance between the carbon atom and the 
center of mass of two oxygen atoms, $r$ is the O$-$O distance, and 
$\gamma$ is  the angle 
between the vectors ${\bf R}$ and ${\bf r}$.
The  calculations are performed using the
expansion of the propagator in terms of Chebyshev polynomials, as implemented
in the program package \lq PolyWave'.\cite{NOTE-CO21-4} 
The settings for the iterative scheme are described in
Refs.\ \onlinecite{G13B} and \onlinecite{G16A}. The coordinate 
discrete variable representation (DVR) 
grids in the calculations are as follows. For 
CO$_2(72^\circ|\tilde{X}^1\!A')$: 160  
potential optimized DVR points in $R$ and $r$, and 200 Gauss-Legendre DVR
points in $\gamma$. For all other bent isomers:  
70 and 110
potential optimized DVR points in $R$ and $r$, respectively, 
and 100 Gauss-Legendre DVR
points in $\gamma$. 
The assigned vibrational states for all isomers with 
the isotopic composition $^{12}$C$^{16}$O$_2$ are given in Table \ref{table2}. 
The same settings are used in the calculations of the vibrational states in 
the isotopically substituted carbon dioxide. 

\begin{acknowledgments}
Financial support by the  Deutsche
  Forschungsgemeinschaft is gratefully acknowledged. 
\end{acknowledgments}


\clearpage
\newpage

\begin{sidewaystable}
\caption{Properties of the bent neutral $C_{2v}$ symmetric 
CO$_2$ isomers: Equilibrium bond length $R_e$; the quantum 
mechanical fundamental vibrational
frequencies (comparisons to other calculations and/or experimental
values are given parenthetically 
where available); depth $\Delta E_{\rm well}$ 
of the 
potential wells of the isomers;  the dipole moments
$\mu_{\rm eq}$ at equilibrium geometries; the Herzberg-Teller coefficients
$|\partial \mu_i/\partial q_n|$ for the components
$\mu_{i=y,z}$ and the symmetry coordinates
$q_n = R_+,\alpha_{\rm OCO}$ and $R_-$; the calculated
intensities $I_{\bm n}$ of the transitions from $(0,0,0)$ state  to
the vibrational states $(1,0,0)$ and $(0,0,1)$ relative to the intensity
$I_{010}$ of the $(0,1,0)$ fundamental. 
The same properties are also shown for 
linear CO$_2$ and the CO$_2^-$ anion radical where available. 
}
\label{table1}
\begin{ruledtabular}
\begin{tabular}{lccccccc}
Isomer & 
CO$_2(70^\circ|X^1\!A')$ & CO$_2(118^\circ|2^1\!A')$ &  
CO$_2(119^\circ|1^3\!A')$ &  CO$_2(127^\circ|1^1\!A'')$ & 
CO$_2(128^\circ|1^3\!A'')$ &   CO$_2^-(138^\circ|1^2\!A')$ 
& CO$_2(180^\circ|X^1\!A')$ \\  
\hline
$R_e$, $a_0$ 
      & 2.51 & 2.36 & 2.36 & 2.37 & 2.37 & 2.33 [\onlinecite{SSHS99}] & 2.20 \\
$\omega_s$, cm$^{-1}$ & 1495 (1480\,[\onlinecite{HM00A}])&  
             1336 (1297\,[\onlinecite{SFCCRWB92}])& 
             1378 (1322\,[\onlinecite{SFCCRWB92}])& 
             1282 (1264\,[\onlinecite{SFCCRWB92}])& 
             1088 & 650 [\onlinecite{SSHS99}] & 1388 \\
$\omega_b$, cm$^{-1}$  & 666 (802\,[\onlinecite{HM00A}])&  
             576 (593\,[\onlinecite{SFCCRWB92}])& 
             602 (591\,[\onlinecite{SFCCRWB92}])& 
             671 (671\,[\onlinecite{SFCCRWB92}])& 
             651 & 1172 [\onlinecite{SSHS99}] & 667 \\
$\omega_a$, cm$^{-1}$  & 683 (830\,[\onlinecite{HM00A}])&  
             846 (758\,[\onlinecite{SFCCRWB92}])& 
             1095 (812\,[\onlinecite{SFCCRWB92}])& 
             904 (789\,[\onlinecite{SFCCRWB92}])& 
             614 & 1657---1690 [\onlinecite{HH66,ZA99,SSHS99}] & 2349 \\
$\Delta E_{\rm well}$, eV & 0.45\tablenotemark[1] &
                        0.25\tablenotemark[2]\tablenotemark[3] &
                        1.2\tablenotemark[2] &
                        0.15\tablenotemark[2]\tablenotemark[3] &
                        0.56\tablenotemark[2] &
                        0.26\tablenotemark[4] &
                        5.45\tablenotemark[2] \\
$\mu_{\rm eq}$, D 
       & -0.25 & 0.99 & 0.46 & 0.36 & 0.25 & 0.27 [\onlinecite{GBC98}] & 0.0 \\
$|{\partial\mu_z}/{\partial R_+}|$, D/$a_0$ 
       & 0.3 & 0.1 & 0.05 & 0.12 & 0.18 &  & \\
$|{\partial\mu_z}/{\partial \alpha}|$, D/rad 
       & 0.15 & 0.1 & 0.15 & 0.20 & 0.25 &  & \\
$|{\partial\mu_y}/{\partial R_-}|$, D/$a_0$ 
       & 1.2 & 0.8 & 1.1 & 0.85 & 1.0 &  & \\
$I_{100}/I_{010}$ 
       & 7.0 & 1.0 & 0.8 & 1.4 & 1.1 &  1.9 [\onlinecite{ZA99}]  & 0 \\
$I_{001}/I_{010}$ 
       & 30 & 107 & 49 & 22 & 9 & 28  [\onlinecite{ZA99}]  & 
12 [\onlinecite{RY81}] 
 \end{tabular}
\end{ruledtabular}
\tablenotetext[1]{Relative to the barrier towards linear OCO minimum.}
\tablenotetext[2]{Relative to the
triplet dissociation threshold 
${\rm CO}(\tilde{X}^1\Sigma^+)+{\rm O}(^3P)$.}
\tablenotetext[3]{The well depth relative to the
singlet dissociation threshold 
${\rm CO}(\tilde{X}^1\Sigma^+)+{\rm O}(^1D)$
is $\sim 2$\,eV.}
\tablenotetext[4]{Relative to the threshold CO$_2 + e$ [\onlinecite{SSHS99}].} 
\end{sidewaystable}

\begin{table}
\caption{Calculated vibrational energies (relative to the 
vibrational ground state, in cm$^{-1}$) for the bent 
singlet and triplet isomers of CO$_2$.  
}
\label{table2}
\begin{ruledtabular}
\begin{tabular}{ccccccc}
No. & $(v_s,v_b,v_a)^a$ & CO$_2(72^\circ|\tilde{X}^1\!A')$& CO$_2(118^\circ|2^1\!A')$& 
CO$_2(119^\circ|1^3\!A')$& CO$_2(127^\circ|1^1\!A'')$& CO$_2(128^\circ|1^3\!A'')$ \\
\hline
   1 & (0,0,0) &        0.0 &        0.0 &          0.0 &      0.0 &        0.0 \\ 
   2 & (0,1,0) &      666.2 &      576.8 &        602.6 &    671.1 &      651.8 \\ 
   3 & (0,0,1) &      682.8 &      846.1 &       1095.1 &    904.7 &      614.7 \\ 
   4 & (1,0,0) &     1494.6 &     1336.1 &       1377.8 &   1282.0 &     1088.4 \\ 
   5 & (0,2,0) &     1362.6 &     1157.3 &       1205.5 &   1341.5 &     1302.0 \\ 
   6 & (0,1,1) &     1321.3 &     1404.4 &       1679.1 &   1559.5 &     1239.6 \\ 
   7 & (0,0,2) &     1268.6 &     1816.0 &       2189.9 &   1899.1 &            \\ 
   8 & (1,1,0) &     2145.9 &     1905.0 &       1970.6 &   1947.9 &     1718.5 \\ 
   9 & (0,3,0) &     2041.7 &     1743.4 &       1809.4 &   2011.8 &            \\ 
  10 & (1,0,1) &     2228.3 &     2114.6 &       2406.2 &   2127.3 &     1622.8 \\ 
  11 & (0,2,1) &     1956.9 &     1975.0 &       2263.1 &   2213.8 &     1861.9 \\ 
  12 & (0,1,2) &     1905.9 &     2370.2 &       2764.5 &   2537.3 &            \\ 
  13 & (2,0,0) &            &     2655.3 &       2743.4 &   2560.2 &            \\ 
  14 & (1,2,0) &            &     2480.7 &       2563.1 &   2612.2 &            \\ 
  15 & (0,4,0) &            &     2333.7 &       2415.0 &   2682.3 &            \\ 
  16 & (1,1,1) &            &     2670.4 &       2977.0 &   2771.7 &     2209.3 \\ 
  17 & (0,3,1) &            &     2550.6 &       2847.6 &   2867.4 &     2481.5 \\ 
  18 & (0,0,3) &     1866.8 &     2798.4 &       3283.8 &   2927.0 &     2632.7 \\ 
  19 & (1,0,2) &            &     3068.3 &       3441.6 &   3086.9 &            \\ 
  20 & (0,2,2) &            &     2939.8 &       3342.8 &   3178.9 &            \\ 
  21 & (2,1,0) &            &     3218.0 &       3317.9 &   3217.2 &            \\ 
  22 & (1,3,0) &            &     3058.5 &       3155.4 &   3276.2 &            \\ 
  23 & (0,5,0) &            &     2922.6 &       3022.3 &   3352.2 &            \\ 
  24 & (2,0,1) &            &     3386.5 &       3715.1 &   3353.0 &            \\ 
  25 & (1,2,1) &            &     3231.1 &       3547.5 &   3416.1 &     2797.8 \\ 
  26 & (0,4,1) &            &     3128.1 &       3433.1 &   3520.6 &            \\ 
  27 & (0,1,3) &            &     3340.3 &              &   3559.0 &            \\ 
  28 & (1,1,2) &            &     3617.5 &              &   3718.1 &            \\ 
  29 & (0,3,2) &            &     3497.7 &              &   3820.7 &            \\ 
  30 & (3,0,0) &            &     3958.5 &              &   3826.1 &           
\end{tabular}
\end{ruledtabular}
\end{table}

\setcounter{table}{1}
\begin{table}
\caption{{\it Continued from previous page}}
\begin{ruledtabular}
\begin{tabular}{ccccccc}
No. & $(v_s,v_b,v_a)^a$ & CO$_2(72^\circ|\tilde{X}^1\!A')$& CO$_2(118^\circ|2^1\!A')$& 
CO$_2(119^\circ|1^3\!A')$& CO$_2(127^\circ|1^1\!A'')$& CO$_2(128^\circ|1^3\!A'')$ \\
\hline
  31 & (2,2,0) &            &     3786.7 &       3888.6 &   3872.7 &            \\ 
  32 & (1,4,0) &            &            &       3746.9 &   3939.3 &            \\ 
  33 & (0,0,4) &            &            &              &   3971.7 &            \\ 
  34 & (2,1,1) &            &            &              &   3985.9 &            \\ 
  35 & (0,6,0) &            &            &       3631.0 &   4021.4 &            \\ 
  36 & (1,3,1) &            &            &              &   4058.3 &            \\ 
  37 & (1,0,3) &            &            &              &   4065.4 &            \\ 
  38 & (0,5,1) &            &            &       4019.2 &   4172.4 &            \\ 
  39 & (0,2,3) &            &            &              &   4194.2 &            \\ 
  40 & (2,0,2) &            &            &              &   4299.9 &            \\ 
  41 & (1,2,2) &            &            &              &   4350.1 &            \\ 
  42 & (0,4,2) &            &            &              &   4462.5 &            \\ 
  43 & (3,1,0) &            &            &              &   4474.3 &            
\end{tabular}
\end{ruledtabular}
\tablenotetext[1]{$v_s$, $v_b$, and $v_a$ are quantum numbers of the symmetric stretch, bend, 
and antisymmetric stretch vibrations.}
\end{table}

\clearpage
\newpage

\begin{figure}[t]
\caption{
(a)  Experimental absorption cross
section of CO$_2$ ($T=195$\,K) as a function of photon energy.\cite{YESPIM96} 
Thin sticks mark the major
experimental peaks and their calculated counterparts (resonance positions). 
Theoretical electronic assignments of major spectral bands\cite{G13B} are 
given in terms of equilibrium 
bending angles and main molecular motions (shown as cartoons). 
(b) Potential energy profiles along the minimum energy path shown as functions
of the    OCO bond angle for the ground electronic state (black), the two 
singlet  states $2^1\!A'$ and $1^1\!A''$ (red) and 
the two triplet states $1^3\!A'$ and $1^3\!A''$  
(blue). Dipole moment vectors are sketched to scale
for each bent equilibrium. The drawings are ordered from left
to right according to the equilibrium angle they illustrate. 
Dashed horizontal lines mark the
positions of the dissociation thresholds 
${\rm CO}(\tilde{X}^1\Sigma^+) + {\rm O}(^1\!D)$ (red; ${\rm D_0(S)}$) and 
${\rm CO}(\tilde{X}^1\Sigma^+) + {\rm O}(^3\!P)$
(blue; ${\rm D_0(T)}$). 
}
\label{fig1}
\end{figure}

\begin{figure}[t]
  \caption{Dipole moment surfaces and infrared spectra of
singlet bent isomers 
CO$_2(70^\circ|X^1\!A')$ (a,d), CO$_2(118^\circ|2^1\!A')$ (b,e), and 
CO$_2(127^\circ|1^1\!A'')$ (c,f). In (a---c), the 
functions $\mu_y$ and $\mu_z$ are shown in the plane $(R_1,R_2)$, with the 
the bending angle fixed  at the respective equilibrium values. 
The antisymmetric
component $\mu_y$ is shown above the diagonal. For $\mu_y$, the dashed (dotted)
contour corresponds to -0.10\,D (-0.18\,D) in (a) and (b), and 0.10\,D
(0.02\,D) in (c). 
The symmetric component $\mu_z$ is shown below the diagonal. For  $\mu_z$,
the dashed (dotted)
contour corresponds to -0.22\,D (-0.38\,D) in (a);  1.07\,D (0.99\,D) in (b);  
and 0.51\,D (0.42\,D) in (c). Black dots indicate equilibrium bond distances. 
In (d---f), vibrational excitations via $\mu_y$ ($\mu_z$)
are shown with blue (red) sticks.
Assignments of the main IR peaks in terms of three vibrational quantum 
numbers $(v_s,v_b,v_a)$ are given. The IR 
intensities are normalized to the intensity of the bending 
fundamental $(0,1,0)$.  
}
\label{ir01}
\end{figure}

\begin{figure}[t]
  \caption{Dipole moment surfaces and infrared spectra of bent isomers 
CO$_2(119^\circ|1^3\!A')$ (a,c), and 
CO$_2(128^\circ|1^3\!A'')$ (b,d). The layout is the same as in Fig.\ 
\ref{ir01}. For $\mu_y$ (above the diagonal), the dashed (dotted)
contour corresponds to 0.26\,D (0.18\,D) in (a) and (b).
For $\mu_z$ (below the diagonal),  
the dashed (dotted)
contour corresponds to 0.51\,D (0.42\,D) in (a) and
0.26\,D (0.18\,D) in (b).  
}
\label{ir02}
\end{figure}

\begin{figure}[t]
\caption{Isotopic shifts relative to $^{12}$C$^{16}$O$_2$ 
in the bending (left panel) 
and the antisymmetric stretch (right panel) progressions of singlet bent
isomers. Different colors distinguish different bent structures;
different symbols distinguish different isotopic substitutions as indicated
in the legends in the two panels. For CO$_2(118^\circ|2^1\!A')$, states 
belonging to the progression $(0,v_b,1)$ are used. 
}
\label{isofig}
\end{figure}

\begin{figure}[t]
\caption{The dependence of the zero-point energies of the bent 
states $2^1\!A'$
(blue) and $1^3\!A'$ (red) on the distance to the image plane of Cu(100)
as predicted by Eq.\ (\ref{diposhift}). Energies are measured relative to
the zero-point vibrational state of CO$_2$ in the linear ground electronic
state (horizontal dashed line). The image plane of Cu(100) (vertical
dashed line) is located $2.3\,a_0$ outside the last Cu plane.  
}
\label{meco2}
\end{figure}


\clearpage
\newpage
\mbox{ }
\vspace{1cm}

\includegraphics[angle=0,scale=0.9]{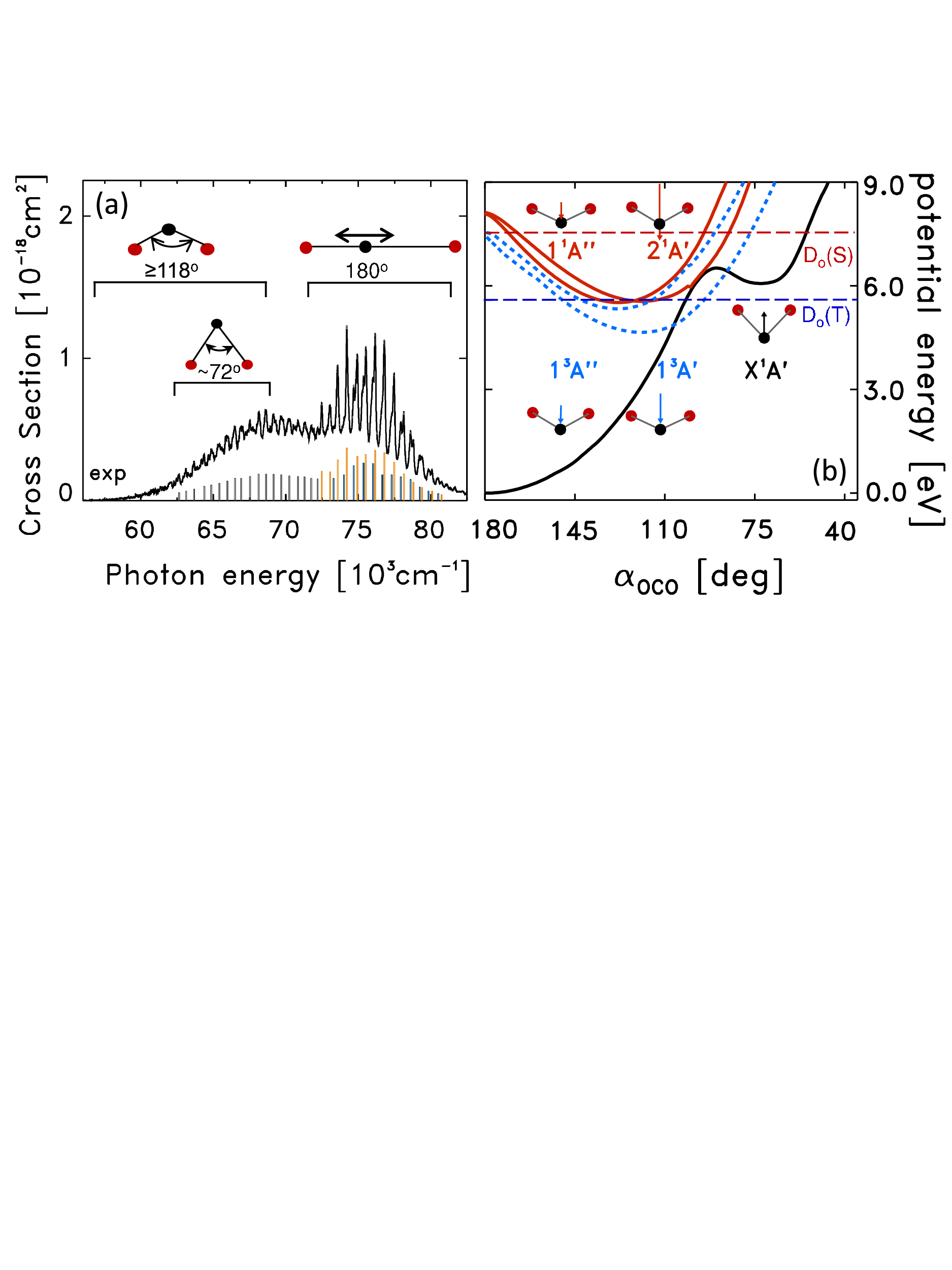}

\vspace{4cm}

Fig.\ 1

\newpage
\mbox{ }
\vspace{1cm}

\includegraphics[angle=0,scale=0.9]{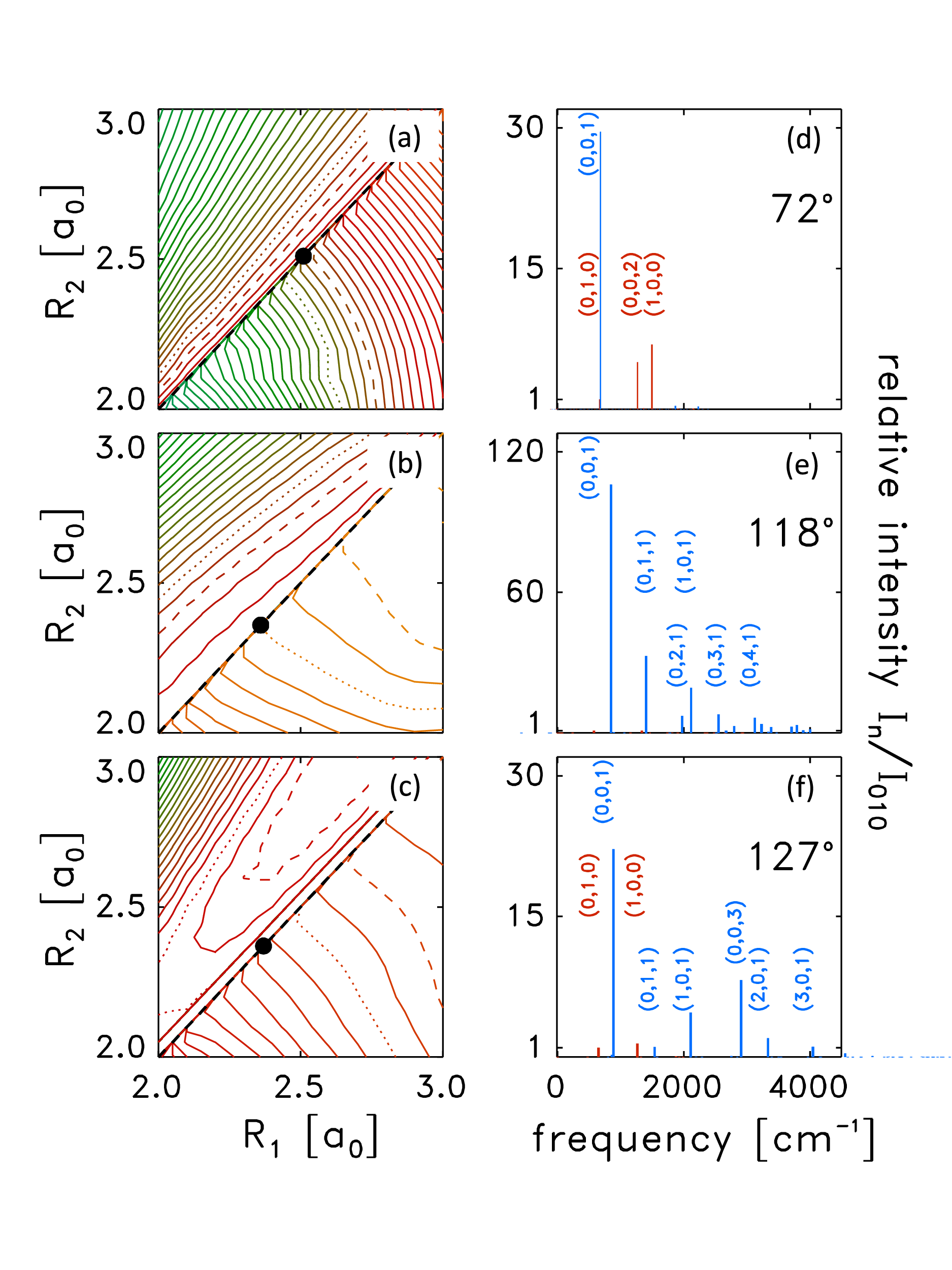}

\vspace{3cm}

Fig.\ 2

\newpage
\mbox{ }
\vspace{1cm}

\includegraphics[angle=0,scale=0.9]{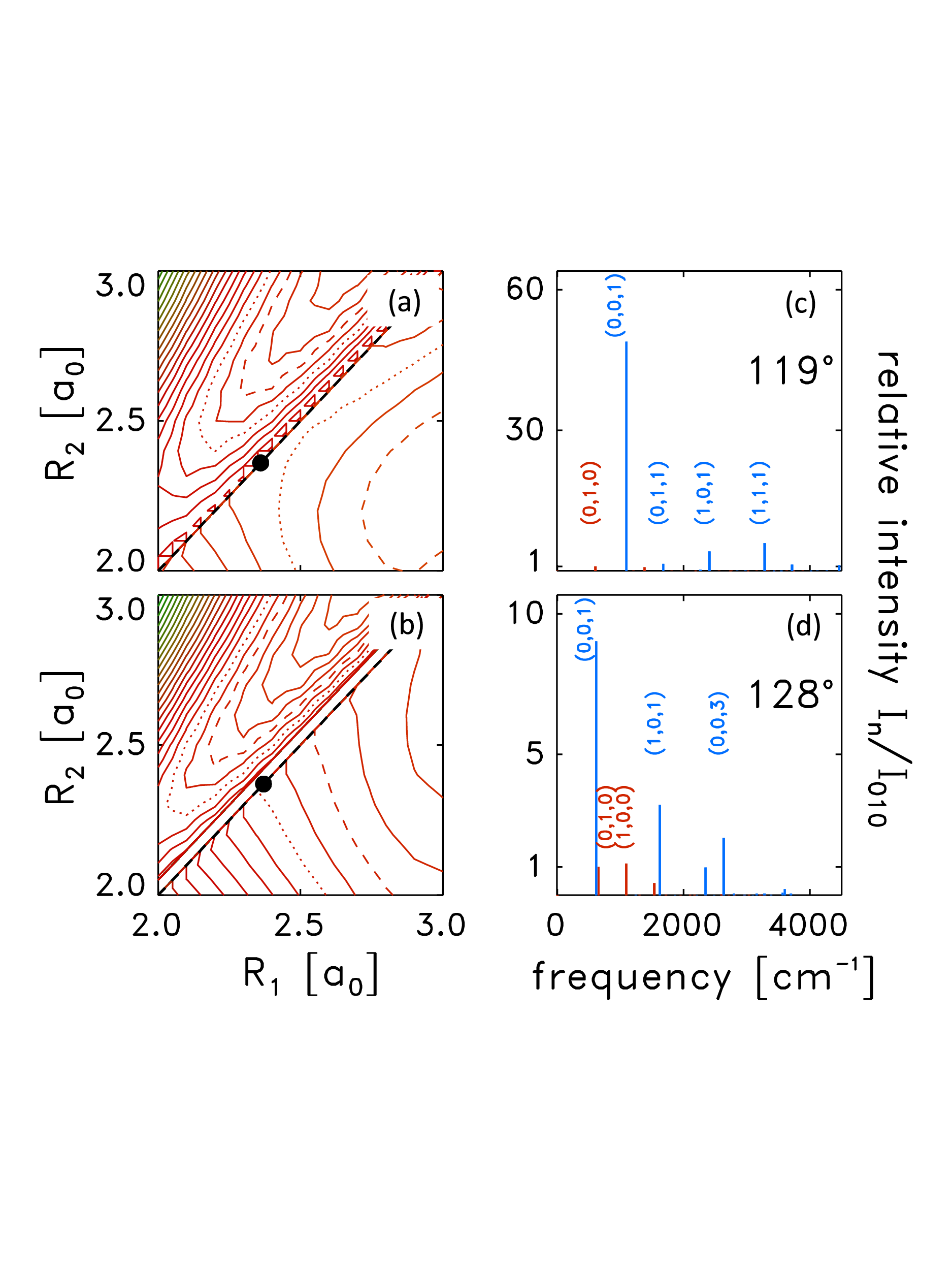}

\vspace{3cm}

Fig.\ 3

\newpage
\mbox{ }
\vspace{1cm}

\includegraphics[angle=90,scale=0.7]{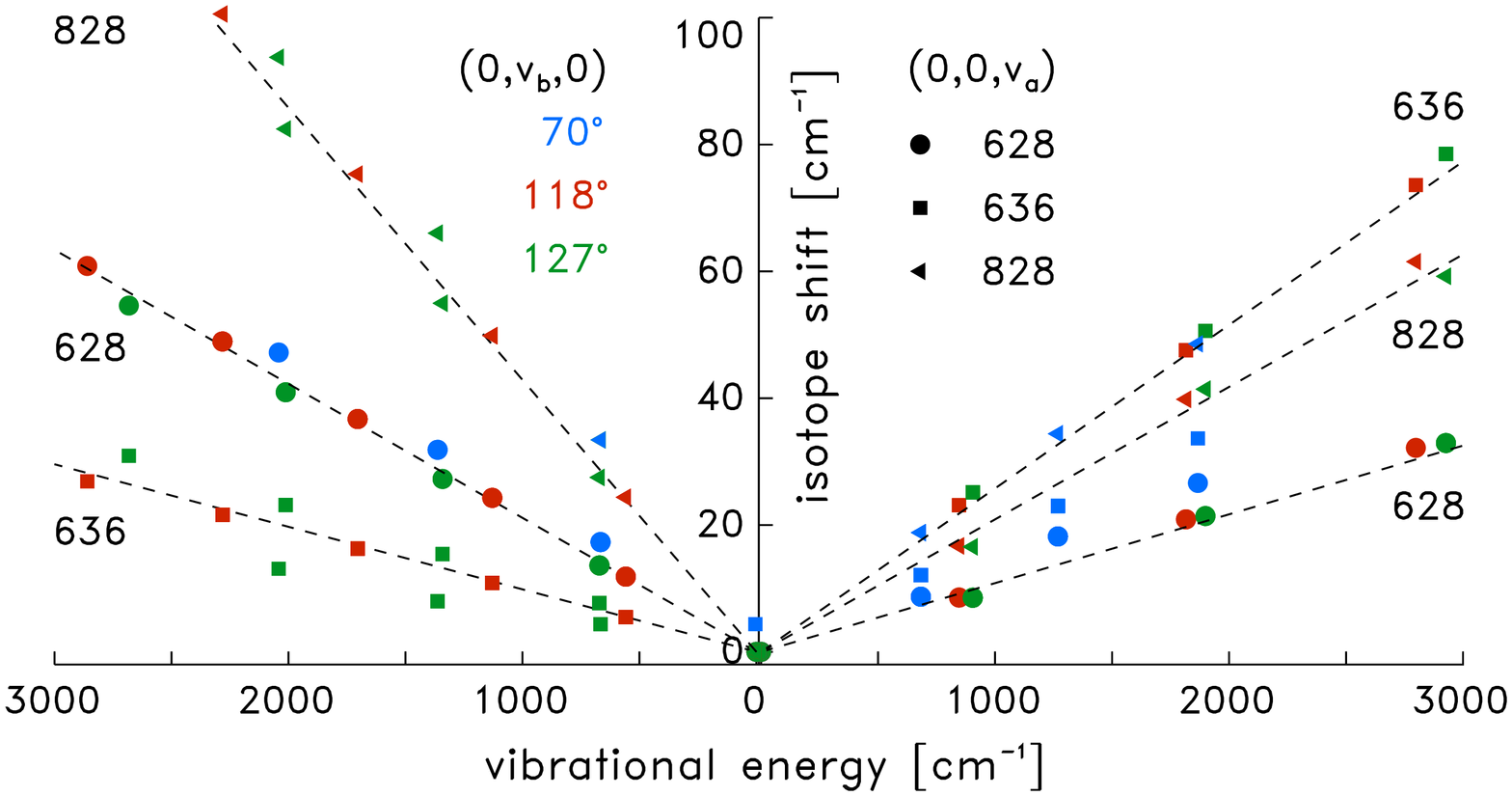}

\vspace{2cm}

Fig.\ 4

\newpage
\mbox{ }
\vspace{1cm}

\includegraphics[angle=0,scale=0.8]{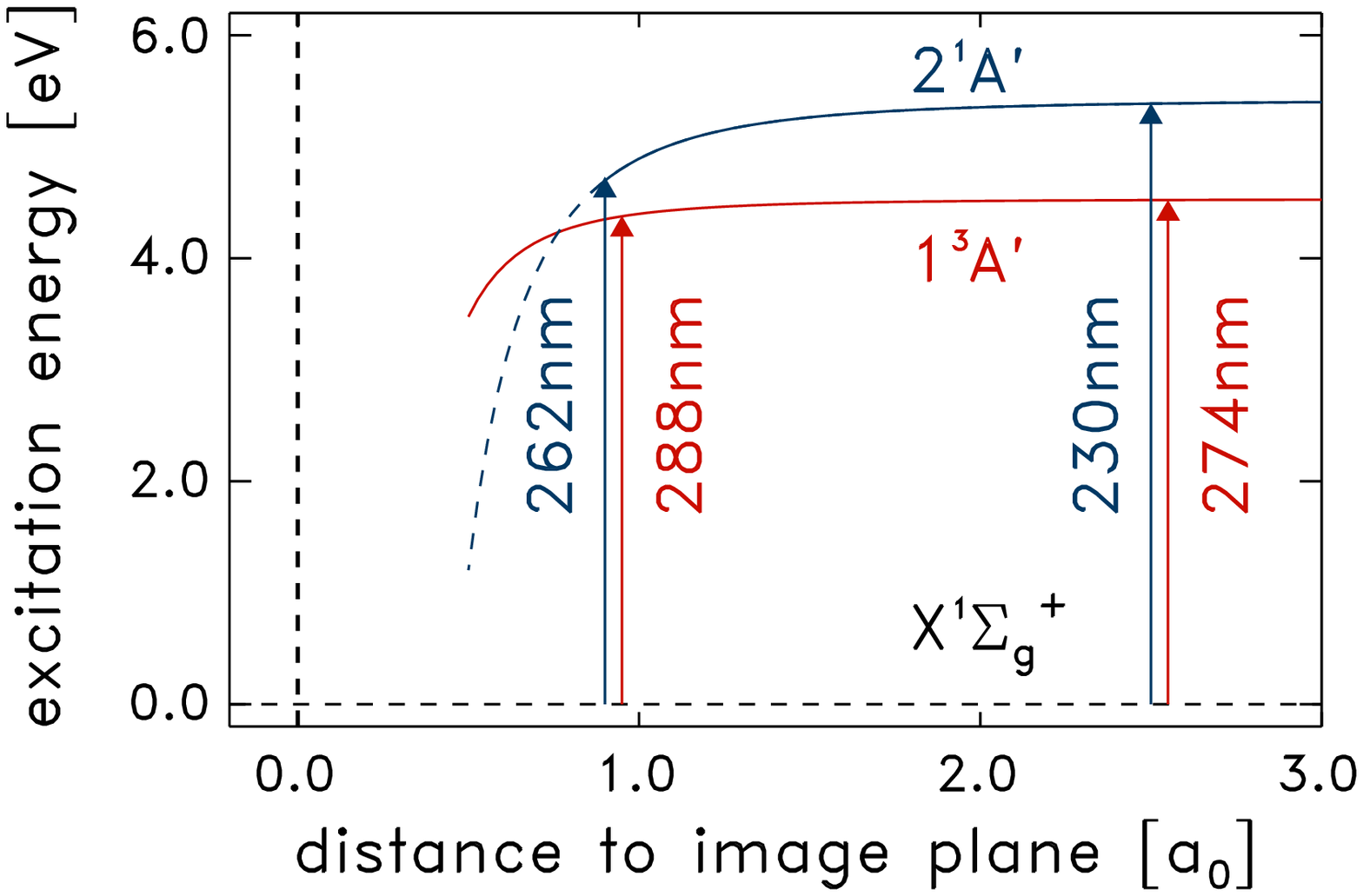}

\vspace{1cm}

Fig.\ 5

\end{document}